\documentclass{aa}

\usepackage{import}
\usepackage{booktabs}
\usepackage{longtable}
\usepackage{graphicx}
\usepackage[varg]{txfonts}
\usepackage{hyperref}
\usepackage{siunitx}
\usepackage{float}
\usepackage{orcidlink}

\usepackage{mathtools}
\usepackage{textgreek}
\usepackage{supertabular}
\usepackage{multirow}

\bibpunct{(}{)}{;}{a}{}{,}

\makeatletter
\renewcommand*\aa@pageof{, page \thepage{} of \pageref*{LastPage}}
\makeatother

\hypersetup{
    colorlinks=true,
    linkcolor=blue,
    filecolor=magenta,      
    urlcolor=blue,
    citecolor=blue,
    pdfpagemode=FullScreen,
}

\begin{document} 

\defcitealias{DAddona2024}{{D24}}

\title{Is there a tilt in the fundamental (hyper)plane?}
\titlerunning{Is there a tilt in the Fundamental (hyper)Plane?}

\authorrunning{D'Addona et al. 2025}
\author{
    M.~D'Addona~\inst{\ref{unisa},\ref{oacn},\ref{unife}}\orcidlink{0000-0003-3445-0483}\and
    A.~Mercurio~\inst{\ref{unisa},\ref{oacn},\ref{infn-unisa}}\orcidlink{0000-0001-9261-7849}\and
    C.~Grillo~\inst{\ref{unimi},\ref{inafmilano}}\orcidlink{0000-0002-5926-7143}\and
    P.~Rosati~\inst{\ref{unife},\ref{oas}}\orcidlink{0000-0002-6813-0632}\and
    G.~Granata~\inst{\ref{unife},\ref{unimi},\ref{portsmouth}}\orcidlink{0000-0002-9512-3788}\and
    G.~Angora~\inst{\ref{oacn},\ref{unife}}\orcidlink{0000-0002-0316-6562}\and
    M.~Annunziatella~\inst{\ref{madrid}}\orcidlink{0000-0002-8053-8040}\and
    P.~Bergamini~\inst{\ref{unimi}, \ref{oas}}\orcidlink{0000-0003-1383-9414}\and
    V.~Bozza~\inst{\ref{unisa},\ref{oacn},\ref{infn-unisa}}\orcidlink{0000-0003-4590-0136}\and
    G.~B.~Caminha~\inst{\ref{max_plank}}\orcidlink{0000-0001-6052-3274}\and
    A.~Gargiulo~\inst{\ref{inafmilano}}\orcidlink{0000-0002-3351-1216}\and
    F.~Getman~\inst{\ref{oacn}\orcidlink{0000-0003-1550-0182}}
    M.~Girardi~\inst{\ref{unitr},\ref{inafts}}\orcidlink{0000-0003-1861-1865}\and 
    A.~Grado~\inst{\ref{oacn}}\orcidlink{0000-0002-0501-8256}\and
    L.~Limatola~\inst{\ref{oacn}}\orcidlink{0000-0002-1896-8605}\and
    M.~Lombardi~\inst{\ref{unimi},\ref{oas}}\orcidlink{0000-0002-3336-4965}\and
    M.~Meneghetti~\inst{\ref{oas}}\orcidlink{0000-0003-1225-7084}\and
    L.~Pecoraro~\inst{\ref{unisa},\ref{oacn}}\orcidlink{0009-0007-9726-2646}\and 
    R.~Ragusa~\inst{\ref{oacn}}\orcidlink{0009-0000-6680-523X}\and
    L. Tortorelli\inst{\ref{munich},\ref{zurich}}\orcidlink{0000-0002-8012-7495}\and
    E.~Vanzella~\inst{\ref{oas}}\orcidlink{0000-0002-5057-135X}
}

\institute{
    Università di Salerno, Dipartimento di Fisica "E.R. Caianiello",
    Via Giovanni Paolo II 132, I-84084 Fisciano (SA), Italy \label{unisa} %
    \and
    INAF -- Osservatorio Astronomico di Capodimonte, Via Moiariello 16, I-80131 Napoli, Italy \label{oacn} %
    \and
    Dipartimento di Fisica e Scienze della Terra, Universit\`a degli Studi di Ferrara, via Saragat 1, I-44122 Ferrara, Italy \label{unife}
    \and
    INFN -- Gruppo Collegato di Salerno - Sezione di Napoli,  Dipartimento di Fisica "E.R. Caianiello", Università di Salerno, via Giovanni Paolo II, 132 - I-84084 Fisciano (SA), Italy. \label{infn-unisa} %
    \and
    Dipartimento di Fisica, Universit\`a  degli Studi di Milano, via Celoria 16, I-20133 Milano, Italy \label{unimi}
    \and
    INAF -- IASF Milano, via A. Corti 12, I-20133 Milano, Italy \label{inafmilano}
    \and
    INAF -- OAS, Osservatorio di Astrofisica e Scienza dello Spazio di Bologna, via Gobetti 93/3, I-40129 Bologna, Italy\label{oas}
    \and
    Institute of Cosmology and Gravitation, University of Portsmouth, Burnaby Rd, Portsmouth PO1 3FX, UK \label{portsmouth}
    \and
    Centro de Astrobiología (CAB), CSIC-INTA, Ctra. de Ajalvir km 4, Torrejón de Ardoz 28850, Madrid, Spain \label{madrid} %
    \and
    Max-Planck-Institut f\"ur Astrophysik, Karl-Schwarzschild-Str. 1, D-85748 Garching, Germany \label{max_plank}
    \and
    Universit\`a degli Studi di Trieste, Dipartimento di Fisica - Sezione di Astronomia, via Tiepolo 11, I-34143 Trieste, Italy \label{unitr}
    \and
    INAF -- Osservatorio Astronomico di Trieste, via G. B. Tiepolo 11, I-34143, Trieste, Italy \label{inafts}
    \and
    Faculty of Physics, University Observatory, Ludwig-Maximilians-Universität München, Munich, Germany \label{munich}
    \and
    Institute for Particle Physics and Astrophysics, ETH Zürich, Zürich, Switzerland \label{zurich}
}

\date{Received 5 March 2025; accepted 24 June 2025}

\abstract
{}
{We investigate the fundamental plane (FP) of selected early-type member galaxies of the galaxy cluster PLCK G287.0+32.9 (\( z_c = 0.3833 \)), also exploring 4D hyperplane extensions.}
{We measured early-type galaxy structural parameters and photometry from \textit{Hubble} Space Telescope observations. We used high-quality spectroscopic data from the Multi Unit Spectroscopic Explorer to measure the galaxy central stellar velocity distribution and stellar population properties. With these data, we constructed the FP through a robust fitting procedure and analyzed its tilt and scatter. We then introduced two hyperplane extensions, one including the stellar mass ($M^\star$-HP) and the other the stellar-over-total mass fraction ($f_{\mathrm{e}}^\star$-HP), and compared their coefficients and scatter to those of the FP.} 
{The FP of PLCK G287.0+32.9 is found to have best-fit parameter values consistent with those in the literature ($\alpha = 1.2 \pm 0.1$ and $\beta = -0.75 \pm 0.04$), with an observed scatter of $0.088$ dex. The $f_{\mathrm{e}}^\star$-HP shows no tilt relative to the theoretical plane ($\alpha = 2.0 \pm 0.3$ and $\beta = -1.1 \pm 0.1$), with an observed scatter of $0.036$ dex, and the $M^\star$-HP has an even tighter relation, with an observed scatter of only $0.036$.}
{Our findings support the idea that the FP is a lower-dimensional projection of a more complex hyperplane and confirm that the variations in the dark matter content contribute significantly to the tilt of the FP. Future studies incorporating larger samples of galaxies and additional physical parameters may further refine our understanding of the FP and its higher-dimensional extensions.}


\maketitle

\section{Introduction}

Early-type galaxies (ETGs) follow scaling relations that connect their photometric and kinematic properties (e.g., \citealt{FaberJackson1976,Kormendy1977a,Kormendy1977b,Kormendy1977c,Kormendy1985}). Among them is the fundamental plane (FP; Eq. \ref{eq:fp_classic}), which is a tight observational correlation among the logarithms of three physical properties of these galaxies, namely the effective radius ($R_{\mathrm{e}}$; i.e., the projected radius of the circle that encloses half of the total light), the average surface brightness ($I_{\mathrm{e}}$) within a circle of radius $R_{\mathrm{e}}$, and the value of the stellar velocity dispersion ($\sigma_{\mathrm{e8}}$) within a circle of radius $R_{\mathrm{e}}/8$. \citep{DjorgovskyDavis1987,Dressler1987}:

\begin{equation}
    \log(R_{\mathrm{e}}) = \alpha\log(\sigma_{\mathrm{e8}}) + \beta\log(I_{\mathrm{e}}) + \gamma.
    \label{eq:fp_classic}
\end{equation}\\

This relation is widely used in various contexts, such as strong lensing modeling \citep[e.g.,][]{Grillo2015,Monna2015,Birrer2020,Granata2022}, and its coefficients can be theoretically derived from a set of assumptions, one of which is that the virial theorem applies, so the dynamical mass of a galaxy within $R_{\mathrm{e}}$ is $M_{\mathrm{e}} \propto k_{\mathrm{vir}} R_{\mathrm{e}} \sigma^2$, where $\sigma$ is the central stellar velocity dispersion. Since the average luminosity within $R_{\mathrm{e}}$ is, by definition, $L_{\mathrm{e}} = L/2 = \pi {R_{\mathrm{e}}}^2 I_{\mathrm{e}}$, and assuming that $\sigma_{\mathrm{e8}} \simeq \sigma$, the total mass-to-light ratio ($M/L$) can be expressed as

\begin{equation}
    \frac{M_{\mathrm{e}}}{L_{\mathrm{e}}} = k_{\mathrm{vir}}\frac{\sigma_{\mathrm{e8}}^2}{R_{\mathrm{e}} I_{\mathrm{e}}} \implies R_{\mathrm{e}} = k_{\mathrm{vir}}\frac{L_{\mathrm{e}}}{M^{\star}_{\mathrm{e}}} \frac{M^{\star}_{\mathrm{e}}}{M_{\mathrm{e}}} \sigma_{\mathrm{e8}}^2  I_{\mathrm{e}}^{-1}~~,
    \label{eq:fp_virial}
\end{equation}\\

\noindent{}where $M^{\star}_{\mathrm{e}} = M^{\star} / 2$ is the stellar mass within the projected stellar half-mass radius ($R^{\star}_{\mathrm{e}}$). In general, stellar mass-to-light ratio ($M^\star/L$) gradients due to radial variations in the star formation, stellar age, and dust attenuation have been shown to make half-light radii systematically bigger than half-mass radii. However, for quiescent galaxies, the light-weighted structural properties provide a good proxy of the mass-weighted properties \citep[for example, see][]{DeGraaff2022a}. In this case, we can assume that $R^{\star}_{\mathrm{e}} = R_{\mathrm{e}}$. Therefore, for a homologous sample of quiescent galaxies that have the same $M^\star/L$ and stellar-over-total mass fraction, within a circle of radius $R_{\mathrm{e}}$, $f_{\mathrm{e}}^{\star} = M^{\star}_{\mathrm{e}} / M_{\mathrm{e}}$, Eq. (\ref{eq:fp_virial}) reduces to Eq. (\ref{eq:fp_classic}) with $\alpha=2$ and $\beta=-1$ \citep{Faber1987}. However, observations show that the FP is tilted with respect to the virial plane, with coefficients varying in the ranges $\alpha \in [0.7, 1.5] $ and $\beta \in [-0.9, -0.6]$, and the FP presents an observed scatter of $\Delta \sim 0.1~\mathrm{dex}$ \citep[see, e.g.,][]{Bender1992,SagliaBenderDressler1993,Jorgensen1996,Phare1998,Bernardi2003,Jorgensen2006,VanDeSande2014}.

The origin of the tilt and scatter of the FP has been debated for a long time, and several studies have shown how different factors can contribute to it, such as the presence of rotational motions \citep{Prugnuel1994,Busarello1997,Weiner2006}, the stellar mass range and star formation rate \citep{DOnofrio2017,Khanday2022,Grebol2023,DOnofrio2024}, the presence of minor and major merging events in the galaxy formation histories \citep{Fernandez2011,Yoon2020, Yoon2022, Marsden2022, Kluge2023}, the dark matter content \citep{Borriello2003,Ferrero2021,DeGraaff2022b}, and the feedback of active galactic nuclei that affect the stellar surface density and the $M/L$ \citep{Rosito2021}. Moreover, non-homology caused by a nonconstant stellar initial mass function and the presence of $M^\star/L$ gradients tend to reduce  $R^{\star}_{\mathrm{e}}$, therefore leading to an overestimation of $M^{\star}_{\mathrm{e}}$ and ultimately causing variations in the estimated total $M/L$ \citep{Bernardi2018, Bernardi2023a, Bernardi2023b,DEugenio2021}.\\

In the past, the three main causes of the tilt and scatter were thought to be the non-homology, a nonconstant $M^\star/L$, and a varying $f^\star_{\mathrm{e}}$. There is now a broad consensus that $M/L$ variations contribute significantly to the tilt of the FP and that variations in the stellar populations seem to account for a large fraction of the scatter. For example, \cite{Cappellari2007} analyzed a sample of 25 ETGs at redshift $z \leq 0.01$ from the Spectroscopic Areal Unit for Research on Optical Nebulae (SAURON) survey (\citealt{Bacon2001}) and showed how $M/L$ variations could account for up to 90\% of the tilt of the FP. This observation was also confirmed by \cite{Cappellari2013}, who obtained a very tight relation ($\Delta \sim 0.07~\mathrm{dex}$) in the mass plane constructed by replacing $I_{e}$ with the total mass within $R_{\mathrm{e}}$ estimated using a dynamic analysis of 260 ETGs from $\mathrm{ATLAS}^{\mathrm{3D}}$ ($z \leq 0.01$; \citealt{Cappellari2011}). Subsequent studies that used the mass plane on larger samples of galaxies, such as \cite{Li2018} and \cite{Zhu2024}, who used 2000 and 6000 galaxies from the Mapping Nearby Galaxies at the APO (MaNGA) survey (\citealt{Bundy2015, Westfall2019}), respectively, found similar results. \cite{HydeBernardi2009a, HydeBernardi2009b} and \cite{DeGraaff2021}, using a sample of 50000 ETGs from the Sloan Digital Sky Survey (SDSS; \citealt{Kollmeier2019}) and 1419 ETGs from the Large Early Galaxy Astrophysics Census (LEGA-C) survey (\citealt{VanDerWel2016, Straatman2018}, up to $z \sim 0.8$), respectively, showed how the observed tilt and scatter of the FP reduce significantly when variations in the $M/L$ ratio are taken into account. In particular, they replaced $I_{\mathrm{e}}$ with the stellar surface mass density ${I^\star}_{\mathrm{e}} = I_{\mathrm{e}} \cdot M^\star/L$, leaving non-homology and variations in the dark matter content as reasons for the remaining tilt. Similar results were also found by \cite{Bezanson2013} using a sample of 16598 ETGs from the SDSS. Finally, \cite{Bernardi2020} further improved these results by replacing $R_{\mathrm{e}}$ with $R_{\mathrm{e}}/k_n$, where $k_n$ depends on the shape of the mass profile (or, equivalently, on the S\'ersic index, $n$), and by extending the plane with the axis ratio as a fourth dimension. The relation they obtained has coefficient values of $\alpha = 1.904 \pm 0.033$ and $\beta = -0.776 \pm 0.019$, which are much closer to those of the viral theorem, and a scatter of only $\Delta \sim 0.05~\mathrm{dex}$.\\

Attempts have also been made to extend the FP to higher-dimensional hyperplanes (HPs). 
For example, \cite{Gargiulo2009} were among the first to extend the FP with stellar population parameters (such as the metallicity, age, and \textalpha-element abundance ratio, $\alpha/\mathrm{Fe}$) using a sample of 141 ETGs from the Shapley Optical Survey \citep[$z \sim 0.049$;][]{Mercurio2006}. They found a strong correlation between the scatter and both the age and the $\alpha/\mathrm{Fe}$, achieving a decrease in the intrinsic scatter to a value as low as $\Delta = 0.049$~dex. Similar results were found, for example, by \citet{Magoulas2012} and \cite{Springob2012}, who highlighted how the scatter in the FP is linked to the age, the metallicity, and other properties of the stellar populations. \cite{Dogruel2023,Dogruel2024}, using 2496 galaxies at $z < 0.12$ from the Galaxy And Mass Assembly (GAMA) survey, constructed a mass HP by adding $\log(n)$ and the $g-i$ color to Eq. (\ref{eq:fp_classic}) to track the dependence on the morphology and the $M^\star/L$ ratio. However, even though they found that this HP has a lower scatter than the original FP and is not specific to a certain class of galaxies, the coefficients measured for passive galaxies, $\alpha = 1.071 \pm 0.011$ and $\beta = -0.627 \pm 0.003$, are still not compatible with those predicted by the virial theorem. Similar results were found by \cite{DEugenio2024}, who built three 4D HPs using the mass-weighted age, the $M^\star/L$ ratio, or their empirical $I_{\mathrm{age}}$ as the fourth dimension, exploiting a sample of $\sim 700$ galaxies with $z < 0.1$ from Sydney-AAO Multi-object Integral (SAMI; \citealt{Croom2012}).\\

In this scientific framework, we exploited \textit{Hubble} Space Telescope (HST) observations and high-quality spectroscopic data derived from observations made with the Multi Unit Spectroscopic Explorer (MUSE), described in \citet[hereafter D24]{DAddona2024}, to build the FP of the members of the galaxy cluster PLCK G287.0+32.9 (PLCK-G287 hereafter), a massive post-merger system at redshift $z_c = 0.3833$ \citep{Bagchi2011,Gruen2014,Bonafede2014,Zitrin2017,Finner2017}. We also constructed two 4D HP extensions of the classic FP using stellar masses derived from stellar population analysis.\\

This paper is structured as follows: In Sect. \ref{sec:data} we describe the spectroscopic and photometric data we used, and in Sect. \ref{sec:analysis} we outline their analysis. In Sect. \ref{sec:fp} we illustrate the construction of the FP of PLCK-G287 and its HP extensions. Finally, in Sect. \ref{sec:discussion} we discuss the obtained results. Throughout this work, we assume a \textLambda\ cold dark matter cosmology with $\Omega_{m0} = 0.3$, $\Omega_{\Lambda0} = 0.7$, and $\mathrm{H}_{\mathrm{0}} = 70~\text{km}~\text{s}^{-1}~\text{Mpc}^{-1}$, for which $\qty{1}{\arcsec} \simeq 5.23~\text{kpc}$ at the redshift of the cluster PLCK-G287.

\section{Data}
\label{sec:data}
We used the same photometric and spectroscopic data presented in \citetalias{DAddona2024}, which we summarize here for the sake of clarity.

\subsection{Photometric data}
PLCK-G287 was observed by the HST in the framework of the Reionization Lensing Cluster Survey (RELICS) survey (P.I.: Dan Coe; \citealt{RELICS}, program ID 14096) using both the Advanced Camera for Surveys (ACS) and the Wide Field Camera 3 (WFC3). Previous observations made with HST in cycle 23 (P.I.: Seitz, program ID 14165; \citealt{Seitz2016}) were also integrated into the RELICS data, for a total of three orbits for the ACS filter group and two orbits for the WFC3 one. We used publicly available images in the two resolutions of $\qty{0.03}{\arcsec}$ (30 mas) and $\qty{0.06}{\arcsec}$ (60 mas) per pixel. They cover an area, centered on the brightest cluster galaxy, with a radius of $\sim\qty{1.7}{\arcmin}$ in the optical bands F435W, F475W, F606W, and F814W, and of $\sim\qty{1.0}{\arcmin}$ in the IR bands F105W, F110W, F125W, F140W, and F160W. For the first group of filters, the total exposure time varies from a minimum of~$2125~\text{s}$~for the F435W band to a maximum of~$4680~\text{s}$~for the F814W band, while for the second group it ranges from a minimum of~$711~\text{s}$~for the F125W band to a maximum of~$11447~\text{s}$~for the F110W band. We ran Sextractor \citep{Sextractor} in dual-image mode on the 60 mas images, using the F814W image as the detection image, and extracted the Kron magnitudes for all available photometric bands.

\subsection{Spectroscopic data}

\label{sub:data.redshift}
We used spectroscopic redshift from the catalog published in \citetalias{DAddona2024}. The redshifts were measured on data acquired with MUSE and the Adaptive Optics Facility (AOF) working in wide field mode \citep{MUSE-AOa,MUSE-AOb} on three nights in March-May 2019  (P.I.: A. Mercurio, ESO program 0102.A-0640(A)). We processed and merged the three datacubes produced by these observations following the prescriptions by \cite{Caminha2019}, using the reduction pipeline version 2.8.3 \citep{Weilbacher2020A}. The final datacube has an exposure time on target of 3.1 hours in two pointings and 3.8 hours in the westernmost pointing. It covers the center of the galaxy cluster for a total area of $\sim 3~\text{arcmin}^2$, with a spatial resolution of $\qty{0.2}{\arcsec}$, and with a spectral resolution of $1.25~\text{\AA}~\text{pixel}^{-1}$ in the vacuum wavelength range $4700~\text{\AA}$ to $9350~\text{\AA}$, with a gap between $5805~\text{\AA}$ and $5967~\text{\AA}$ due to the emission generated by the guiding laser of the AOF.\\

\section{Data analysis}
\label{sec:analysis}
We considered only the spectroscopic members, which were defined in \citetalias{DAddona2024} using the redshift range $0.360 \leq z \leq 0.405$. Since the relations that define the FP are only valid for virialized ETGs, we selected a sample of passive ETG candidates by selecting the members that are not contaminated by other very close bright galaxies or image artifacts (like star diffraction spikes), and the spectra of which did not present any emission line. We then fitted their light in the 30 mas HST F814W image with Morphofit\footnote{Morphofit: a Python package, based on Galfit \citep{Peng2002,Peng2010a} and Sextractor \citep{Sextractor}, for the morphological analysis of galaxies. See \url{https://github.com/torluca/morphofit}}~\citep{Morphofit2023, Tortorelli2023}, using a single S\'ersic component model, convolved with the point spread function, and measured their morphological structural parameters from the best fits. We checked the goodness of the fits by inspecting the 2D residual images and the distribution of the residuals and removed from the sample the galaxies that showed evident spiral arms or substructures that could indicate recent merging events or tidal interactions with faint neighbors. For each of the remaining galaxies in the ETG sample, we re-extracted the 1D spectrum with Specex\footnote{python-specex: a Python package that provides a set of functions and scripts to handle spectral datacubes, see: \url{https://github.com/mauritiusdadd/python-specex}.} using a circular aperture with a diameter of $\qty{1.6}{\arcsec}$ and spatially weighting the extraction with its best-fit light model. This has been shown to minimize the effect of possible contamination and maximize the spectrum signal-to-noise ratio \citep[S/N; see, e.g.,][]{Granata2022}, while allowing for line-of-sight stellar velocity dispersion (LOSVD) estimates that are on average equivalent to those within the galaxy effective radius (Granata et al., in prep.). From the spectra, we then measured the LOSVD and the spectrum S/N with pPXF\footnote{pPXF: python package that implements a penalized PiXel-Fitting method (pPXF) to extract the stellar or gas kinematics and stellar population from absorption-line spectra of galaxies. See \url{https://www-astro.physics.ox.ac.uk/~cappellari/software/}}  \citep{ppxf2023,Cappellari2017,CappellariEmsellem2004}, using the same configuration described in \cite{Bergamini2019}. They found that statistical errors are generally underestimated by $\sim20\%$ for S/N $> 15$, and up to $\sim25\%$ for S/N of $\sim10$. Thus, we corrected the uncertainties on the LOSVD accordingly. We built our final sample following the criteria introduced by \cite{Tortorelli2018} and \cite{Bergamini2019} and selected a total of $35$ ETGs with a spectrum $\mathrm{S/N} \geq 10$, $\mathrm{LOSVD} \geq 80~\text{km}~\text{s}^{-1}$, and $n \geq 2.5$ (see Fig. \ref{fig:hists}).\\

Finally, we exploited the software Bagpipes\footnote{Bagpipes: a Python package to build complex model galaxy spectra and fit them to arbitrary combinations of spectroscopic and photometric data through the MultiNest nested sampling algorithm \citep{Feroz2008,Feroz2009,Feroz2019}. See \url{https://github.com/ACCarnall/bagpipes}.} \citep{Carnall2018,Carnall219b} to simultaneously fit the HST multiband photometry and the spectrum of each ETG, and infer their star formation histories and physical parameters, such as the total stellar mass $M^\star$. By default, Bagpipes uses a \cite{Kroupa2002} initial mass function. We modeled the star formation histories as a delayed exponentially decaying function \citep{Carnall2019a} that has a star formation rate described by
\begin{equation}
    \label{eq:sfr_delayed}
    \mathrm{SFR_{del}}(t) \propto \begin{cases}
    (t - T_0) \exp{\left( -\frac{t - T_0}{\tau} \right)} &\mathrm{if}~~t~>~T_0\\
    0 &\mathrm{if}~~t~<~T_0
    \end{cases}
.\end{equation}

We imposed uniform priors on both the time since the star formation began ($T_0$) and the declining timescale ($\tau$), allowing them to vary in the range $0.1$ to $15$ Gyr and from $0.01$ to $10$ Gyr, respectively. We also assumed uniform priors on the metallicity ($Z$) in the range $0$ to $2.5$. We modeled the dust attenuation with the Calzetti law \citep{Calzetti2000,Calzetti2001}, allowing the attenuation parameter ($A_V$) to vary uniformly from $0$ to $8~\mathrm{mag}$. To reduce the number of free parameters, we fixed the redshift and the stellar velocity dispersion to the values we measured from the spectra. The values of the parameters we measured for the 35 ETGs are reported in Table \ref{selected_member_tables}.

\section{The fundamental (hyper)plane of PLCK-G287}
\label{sec:fp}
From the Morphofit effective semimajor axis $a_{\mathrm{e}}^{\mathrm{(gal)}}$\footnote{$a_{\mathrm{e}}^{\mathrm{(gal)}}$ is the semimajor axis of the ellipse that encloses half of the total light of the galfit model of the galaxy.} and axis ratio $q$, we computed the circularized effective radius $R_{\mathrm{e}} = a_{\mathrm{e}}^{\mathrm{(gal)}} \sqrt{q}$. This choice is also supported by the fact that studies based on cosmological simulations, like that of \citet{DeGraaff2022a}, have shown that using the circularized effective radii, instead of the effective semimajor axes, produces FPs that are closer to the ones built using the actual 3D effective radii. We then computed the mean surface brightness $\mu_{\mathrm{e}}$ by dividing the galfit F814W magnitude $m_{\mathrm{F814W}}^{\mathrm{(gal)}}$ by the circularized effective area (Eq. \ref{eq:mue}; \citealt{Dressler1987}), and then converting it trough Eq. (\ref{eq:logIe}) to $I_{\mathrm{e}}$, which is expressed in units of solar luminosity $L_\odot$ per square parsec ($4.52$ is the absolute AB magnitude of the Sun in the HST F814W band, as reported in Table 3 of \citealt{Willmer2018TheFilters}):

\begin{align}
\label{eq:mue}\mu_{\mathrm{e}} &= m_{\mathrm{F814W}}^{\mathrm{(gal)}} + 2.5\log(2\pi) + 5\log\left(\frac{R_{\mathrm{e}}}{\qty{1}{\arcsec}} \right)\\
\label{eq:logIe}\log \left(I_{\mathrm{e}} \right) &= -0.4 \left[ \mu_e - 21.572 - 10\log(1 + z_c) - 4.52 \right]~~.
\end{align}

Since software like Galfit/Morphofit tends to underestimate the uncertainties on the morphological parameters, following the works \cite{Tortorelli2018} and \cite{Granata2022}, we chose $0.1$, $0.09~\mathrm{kpc}$, and $0.08$ as a lower limit for the uncertainties on $m_{\mathrm{F814W}}^{\mathrm{(gal)}}$, $R_{\mathrm{e}}$, and $\log(I_{\mathrm{e}})$, respectively. Abiding to the prescriptions of \cite{Jorgensen1996}, we computed $\sigma_{\mathrm{e8}}$ by correcting the LOSVD to an aperture of $R_{\mathrm{e}} / 8$ using Eq. (\ref{eq:re8}) and Eq. (\ref{eq:alphacirc}) from \cite{Zhu2023}, where we used the value $\kappa = -0.033$ measured by \cite{DeGraaff2021}:

\begin{align}
\label{eq:re8}\sigma_{\mathrm{e8}} &= \mathrm{LOSVD} \cdot \left( \frac{R_{\mathrm{e}} / 8}{R_{\mathrm{e}}}\right)^{\kappa_{circ}}\\
\label{eq:alphacirc}\kappa_{circ} &= \kappa - 0.106(1 - q)^{4.73}~~.
\end{align}

\begin{figure}
    \centering
    \includegraphics[width=1.0\linewidth]{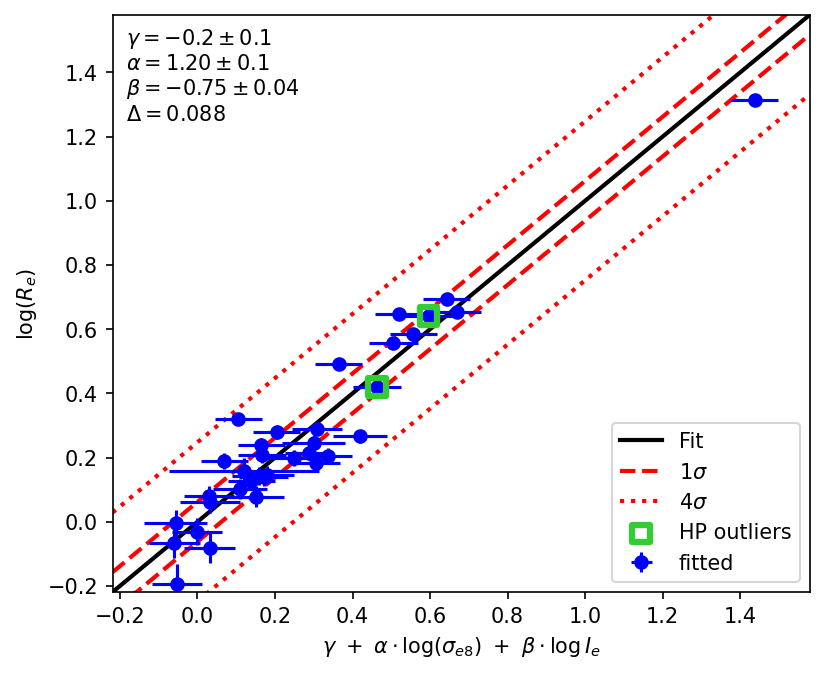}
    \caption{Fit to the classic FP (Eq. \ref{eq:fp_classic}) built with a sample of 35 secure ETGs selected as members of the galaxy cluster PLCK-G287. The aperture-corrected central stellar velocity dispersion ($\sigma_{\mathrm{e8}}$) is expressed in $\mathrm{km}~\mathrm{s}^{-1}$, the surface brightness ($I_{\mathrm{e}}$) in  $L_\odot~\mathrm{pc}^{-2}$, and the circularized effective radius ($R_{\mathrm{e}}$) in kpc. The solid black line is the projection of the best-fitting plane, and the dashed and dotted red lines indicate the $1$ and $4$ standard deviation intervals, respectively. Green squares indicate the two objects that are found to be outliers of the $M^\star$ and $f_{\mathrm{e}}^\star$ HPs.}
    \label{fig:fp_classic}
\end{figure}

We used ltsfit\footnote{ltsfit: a Python package for very robust HP fitting in N dimensions. See \url{https://pypi.org/project/ltsfit}.} \citep{Cappellari2009,Cappellari2013} to fit the FP of PLCK-G287 using $\sigma_{\mathrm{e8}}$ expressed in $\mathrm{km}~\mathrm{s}^{-1}$, $I_{\mathrm{e}}$ in $L_\odot~\mathrm{pc}^{-2}$, and $R_{\mathrm{e}}$ in kpc. Following the ltsfit guidelines, we also increased the $\sigma$-clipping threshold from the default value of $2.6$ standard deviations to 4 standard deviations to reduce the number of objects identified as outliers. The best-fitting plane, shown in Fig. \ref{fig:fp_classic}, has coefficients $\alpha = 1.2 \pm 0.1$ and $\beta = -0.75 \pm 0.04$, and an observed (intrinsic) scatter of $\Delta = 0.088$ ($0.05 \pm 0.02$). These values are consistent with those reported in the literature.\\

\begin{figure*}
    \centering
    \includegraphics[width=0.45\linewidth]{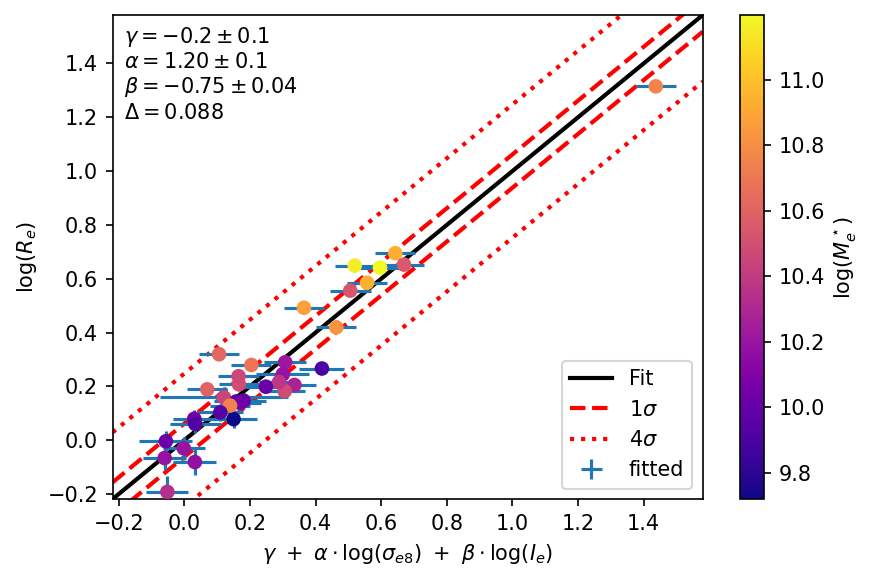}
    \includegraphics[width=0.45\linewidth]{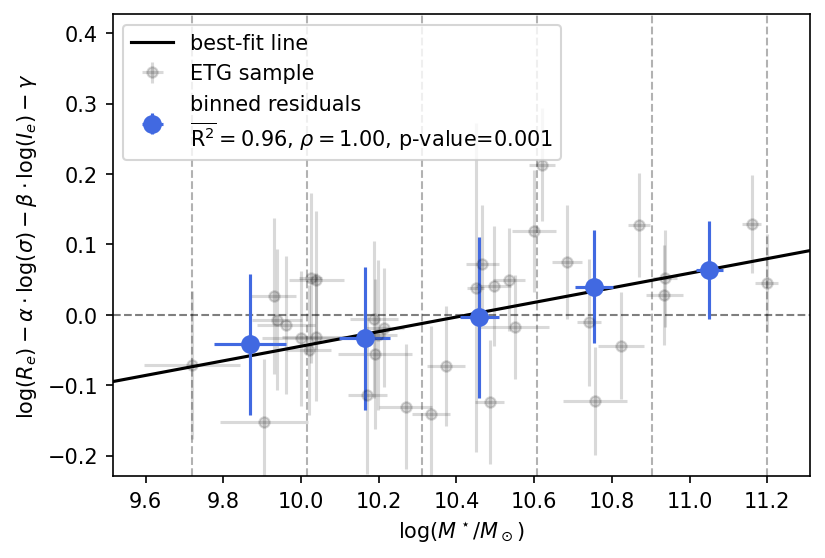}
    \caption{Classic FP of PLCK-G287 with points colored as a function of the stellar mass (left panel) and residuals of the classic FP as a function of $M^\star$ (right panel). Gray points are the ETGs used to fit the FP. A linear fit (solid black line) shows a clear relation between $M^\star$ and the residuals. As a reference, we also show the average residuals (blue points) computed in five bins of equal width of the stellar mass, which are indicated by the vertical dashed gray lines. The value of  $\overline{\mathrm{R}^2} = 0.96$ and the Spearman's rank correlation coefficient $\rho=1$ with a p-value of $0.001$, computed using these binned residuals, confirm the linear relation with the stellar mass.}
    \label{fig:fp_res_mstar}
\end{figure*}

By correlating the residuals of the best-fitting plane with the physical parameters inferred with Bagpipes (such as the metallicity, mass-weighted age, $\tau$, $T_0$, $A_V$, etc.), we found a clear trend only with the stellar mass (Fig. \ref{fig:fp_res_mstar}). We also checked the correlation between FP residual and some empirical properties like the velocity dispersion, the HST magnitudes in the bands F160W and F110W  (which correspond approximately to rest-frame J and I magnitudes), and the HST F606$-$F110W color (which corresponds approximately to a rest-frame $g-i$ color). We find that only the F160W and F110W magnitudes show a clear correlation with the FP residuals, and this is consistent with the fact that FP residuals correlate with the spectral energy distribution-fitting stellar masses, since IR magnitudes are generally good proxies for these masses. We therefore extended the FP by using $\log(M^\star)$ as a fourth dimension ($M^\star$-HP, Eq. \ref{eq:fhp_mstar}), with $M^\star$ expressed in units of $\mathrm{M}_{\mathrm{\odot}}$:

\begin{equation}
    \log(R_{\mathrm{e}}) = \alpha\log(\sigma_{\mathrm{e8}}) + \beta\log(I_{\mathrm{e}}) + \epsilon\log(M^\star) + \gamma~~.
    \label{eq:fhp_mstar}
\end{equation}\\

\begin{figure}
    \centering
    \includegraphics[width=1.0\linewidth]{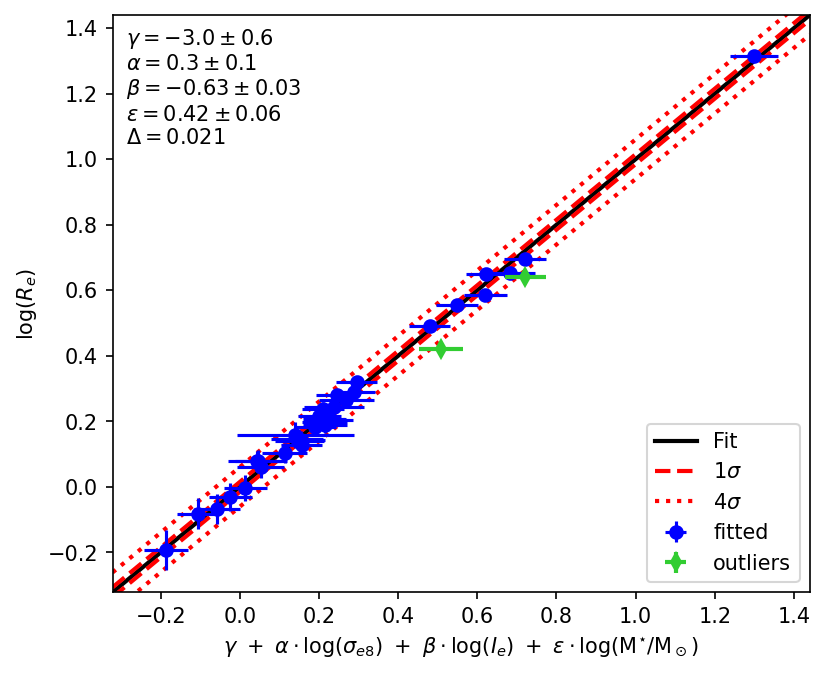}
    \caption{Fit to the stellar mass HP (Eq. \ref{eq:fhp_mstar}) built with a sample of 35 secure ETG members of the galaxy cluster PLCK-G287.\ The aperture-corrected central velocity dispersion ($\sigma_{\mathrm{e8}}$) is expressed in $\mathrm{km}~\mathrm{s}^{-1}$, the surface brightness ($I_{\mathrm{e}}$)  in $L_\odot~\mathrm{pc}^{-2}$, the stellar mass ($M^\star$) in $\mathrm{M}_{\mathrm{\odot}}$, and the circularized effective radius ($R_{\mathrm{e}}$) in kpc. The solid black line is the projection of the best-fitting plane, and the dashed and dotted red lines indicate the $1$ and $4$ standard deviation intervals, respectively. Green points are objects identified as outliers and therefore not included in the fit.}
    \label{fig:fhp_mstar}
\end{figure}

As we did for the classic FP, we fitted this stellar mass HP with ltsfit and the best-fit plane resulted in a very tight relation with an observed scatter of $\Delta=0.023$ (Fig. \ref{fig:fhp_mstar}), with only two ETGs identified as $4\sigma$ outliers. We note that this scaling relation shows an intrinsic scatter that is consistent with zero, which could indicate the presence of some form of correlated noise. The coefficients $\alpha = 0.3 \pm 0.1$, $\beta = -0.63 \pm 0.03$, and $\epsilon = 0.42 \pm 0.06$ cannot be directly linked to those of the classic FP.

We then relaxed the hypothesis that $f^{\star}_{\mathrm{e}}$ is constant across the ETGs sample, while assuming that the virial theorem still holds and that $M_{\mathrm{e}}^\star/L_{\mathrm{e}}$ is constant. In that case, Eq. (\ref{eq:fp_virial}) can be written as Eq. \ref{eq:fhp_fstar} ($f_{\mathrm{e}}^\star$-HP), which describes a 4D stellar mass fraction HP extension of the classical FP, where the coefficients $\alpha$ and $\beta$ theoretically are still $2$ and $-1$, respectively, and $\epsilon = 1$: 

\begin{equation}
    \label{eq:fhp_fstar}
    \log(R_{\mathrm{e}}) = \alpha \log(\sigma_{\mathrm{e8}}) + \beta \log(I_{\mathrm{e}}) + \epsilon \log(f_{\mathrm{e}}^\star) + \gamma~~.
\end{equation}

\begin{table*}[t]
    \centering
    \caption{FP and HP equations with the corresponding best-fitting parameters and figure of merits.}
    \renewcommand{\arraystretch}{1.5}
    \resizebox{\textwidth}{!}{%
    \begin{tabular}{|r|c|c|c|c|c|c||c|c|c|}
       \cline{2-10}
       \multicolumn{1}{c|}{} & \multirow{2}{*}{(Hyper-)plane equation} & \multicolumn{4}{c|}{Best-fit parameters} & Observed scatter &  \multirow{2}{*}{$\Delta\overline{\mathrm{R^2}}$} & \multirow{2}{*}{$\Delta$AIC}  & \multirow{2}{*}{$\Delta$BIC}\\
       \cline{3-6}
       \multicolumn{1}{c|}{} & {} & $\alpha$ & $\beta$ & $\epsilon$ &  $\gamma$ & $\Delta$ & & &\\
       \midrule
       FP & \multicolumn{1}{l|}{$\log(R_{\mathrm{e}}) = \alpha\log(\sigma) + \beta\log(I_{\mathrm{e}}) + \gamma$} & $1.2 \pm 0.1$ & $-0.75 \pm 0.04$ & {} & $-0.2 \pm 0.1$ & $0.088$ & $0$ & $0$ & $0$ \\
       $M^\star$-HP & \multicolumn{1}{l|}{$\log(R_{\mathrm{e}}) = \alpha\log(\sigma_{\mathrm{e8}}) + \beta\log(I_{\mathrm{e}}) + \epsilon\log(M^\star) + \gamma$} & $0.3 \pm 0.1$ & $-0.63 \pm 0.03$ & $0.42 \pm 0.06$ & $-3.0 \pm 0.6$ & $0.021$ & $+0.081$ & $-67.7$ & $-66.6$\\
       $f_{\mathrm{e}}^\star$-HP & \multicolumn{1}{l|}{$\log(R_{\mathrm{e}}) = \alpha \log(\sigma) + \beta \log(I_{\mathrm{e}}) + \epsilon \log(f_{\mathrm{e}}^\star) + \gamma$} & $2.0 \pm 0.3$ & $-1.1 \pm 0.1$ & $0.8 \pm 0.2$ & $-0.68 \pm 0.09$ & $0.036$ & $+0.060$ & $-28.6$ & $-27.6$\\
       \bottomrule
    \end{tabular}
    }
    \label{tab:fit_params}
    \tablefoot{$\Delta\overline{\mathrm{R^2}}$, $\Delta$AIC, and  $\Delta$BIC are the differences relative to the values computed for the FP, which are $\overline{\mathrm{R^2}}=0.908$, $\mathrm{AIC} = -67.3$, and $\mathrm{BIC} = -63.4$. A positive value of $\mathbf{\Delta}\overline{\mathrm{R^2}}$ and a negative value of $\mathbf{\Delta}$AIC and $\mathbf{\Delta}$BIC are considered as a strong indication of a better model.}
\end{table*}

Observations show that the total mass distribution of ETGs is well described by a density profile in the form of $\rho(r) \propto r^{\gamma'}$, with the slope parameter $\gamma'$ varying as a function of the redshift, the total mass, and the local environment. Simulations, in fact, suggest that ETGs had a steeper density slope in the early universe ($\gamma' \sim -3$ at $z \sim 3$), which is flattened by dry merger events \citep[e.g.,][]{Remus2017}. Recent studies indicate that ETGs in the local universe follow a near-isothermal profile, up to $4R_{\mathrm{e}}$, with an average value of $\langle\gamma'\rangle$ varying in the range $-2.0$ to $\sim -2.2$ \citep[see, e.g.,][]{Cappellari2015,Poci2017,Bellstedt2018,Auger2010}. Furthermore, \citet[]{Derkenne2023} has found that ETGs residing in the dense local environment of a cluster at redshift $z=0.348$ have a median density slope of $\langle\gamma'\rangle \sim -2.01 \pm 0.04$, suggesting that cluster environments are more likely to host galaxies with shallower mass distributions than field environments. Density slopes of $\langle\gamma'\rangle = -2$ are called "isothermal" as they correspond to the density slope of a singular isothermal sphere (SIS; see, e.g., \citealt{Mould1990} and \citealt{Arnaboldi1998}):

\begin{equation}
    \label{eq:sis_rho}
    \rho(r) = \frac{\sigma_{\mathrm{SIS}}^2}{2 \pi G r^2}~~.
\end{equation}

Other strong gravitational lensing studies have also shown that the SIS can be used as a good approximation of the total density profile of ETGs within $1R_{\mathrm{e}}$ \citep[for example, ]{Koopmands2006,Koopmans2009,Barnabe2009,Barnabe2010}, and it has been tested that $\sigma_{\mathrm{SIS}} \simeq \sigma_{\mathrm{e8}}$ \citep[e.g.,][]{Treu2006,Bolton2008V,Bolton2008VII,Grillo2008}, the projected total mass $M_{\mathrm{e}}$ enclosed in a cylinder of radius $R_{\mathrm{e}}$ can be derived from

\begin{equation}
    \label{eq:mtot_e}
    M_{\mathrm{e}} = \frac{\pi}{G} \sigma_{\mathrm{e8}}^2 R_{\mathrm{e}}~~.
\end{equation}

\begin{figure}
    \centering
    \includegraphics[width=1.0\linewidth]{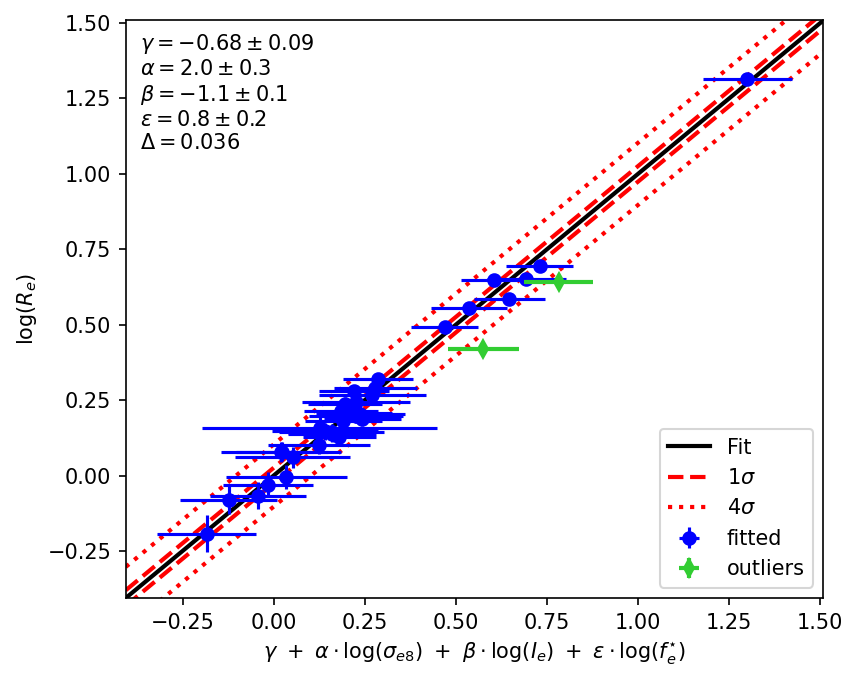}
    \caption{Fit to the stellar mass fraction HP (Eq. \ref{eq:fhp_fstar}) built with a sample of 35 secure ETG members of the galaxy cluster PLCK-G287. The aperture-corrected central velocity dispersion ($\sigma_{\mathrm{e8}}$) is expressed in $\mathrm{km}~\mathrm{s}^{-1}$, the surface brightness ($I_{\mathrm{e}}$) in $L_\odot~\mathrm{pc}^{-2}$, and the circularized effective radius ($R_{\mathrm{e}}$) in kpc. The solid black line is the projection of the best-fitting plane, and the dashed and dotted red lines indicate the $1$ and $4$ standard deviation intervals, respectively. Green points are objects identified as outliers and therefore not included in the fit.}
    \label{fig:fhp_fstar}
\end{figure}

\noindent{}Therefore, $f_{\mathrm{e}}^\star = 0.5M^\star / M_{\mathrm{e}}$ (see Fig. \ref{fig:hists}). The ltsfit best-fitting plane has coefficients $\alpha = 2.0 \pm 0.3$, $\beta = -1.1 \pm 0.1$, and $\epsilon = 0.8 \pm 0.2$, which are compatible with the theoretical ones (Fig. \ref{fig:fhp_fstar}). The observed scatter, $\Delta = 0.042$, is higher than the value for the $M^\star$-HP; however, it is still less than half of the scatter of the FP. As in the previous case, the intrinsic scatter is null, and only two galaxies are identified as $4\sigma$ outliers.\\

Since the intrinsic scatter for the two HPs is consistent with zero, to check whether the reduction of the total observed scatter is an actual effect, for the FP and the two HPs, we also computed the Ezekiel adjusted coefficient of determination \citep[$\overline{\mathrm{R^2}}$;][]{Raju1997}, the Akaike information criterion \citep[AIC;][]{Akaike1992}, and the Bayesian information criterion \citep[BIC;][]{Schwarz2007}. Their values are summarized in Table \ref{tab:fit_params}, along with the best fitting parameters. The $\overline{\mathrm{R^2}}$ extends the classic coefficient of determination by accounting for its tendency to increase when additional variables are added to the model. Like the unadjusted $R^2$, it measures how well the model replicates the observed data, but it increases only when a newly added variable provides meaningful explanatory power. The BIC and the AIC, although similar, try to answer to different questions: the AIC is based on the assumption that all models are just approximations, and it tries to identify the one that best describes an unknown, high-dimensional reality. In contrast, the BIC tries to identify the "true" model that generated the observed data. In practice, it is unlikely that any candidate model perfectly reflects the reality (or that the "true" underlying model is among those tested). Therefore, the best one can do is to select the model that better approximates it: lower values of AIC and BIC indicate better models, relative to those tested, and higher values of $\overline{\mathrm{R^2}}$ indicate a better fit to the observed data. For both $f_{\mathrm{e}}^\star$-HP and $M^\star$-HP, the higher values of $\overline{\mathrm{R^2}}$ together with the lower values of AIC and BIC with respect to those of the FP, indicate that the decrease in the scatter in these two HPs cannot be attributed to, for example, the increase in the dimensionality of the parameter space, but rather to the additional information brought in by the stellar mass values. Finally, it has been shown that the value of the coefficients of the best-fitting (hyper-)planes tends to depend also on the specific model used. Since other studies have shown that the FP can be also modeled as a 3D Gaussian  \citep[see, e.g.,][]{Saglia2001,Magoulas2012}, as a further diagnostic check, we repeated the fits using hyperfit\footnote{HyperFit is a simple python package designed to fit N dimensional data with (possibly covariant) errors with an N-1 dimensional plane. See \url{https://github.com/CullanHowlett/HyperFit}} \citep{Robotham2015}, which also takes into account the covariances among the input data. We find that, for the FP and the two HPs, the best-fitting value of the coefficients and the observed scatter are consistent within $1\sigma$ with the ones found using ltsfit.

\section{Discussion}
\label{sec:discussion}

\begin{figure*}
    \centering
    \includegraphics[width=0.3\linewidth]{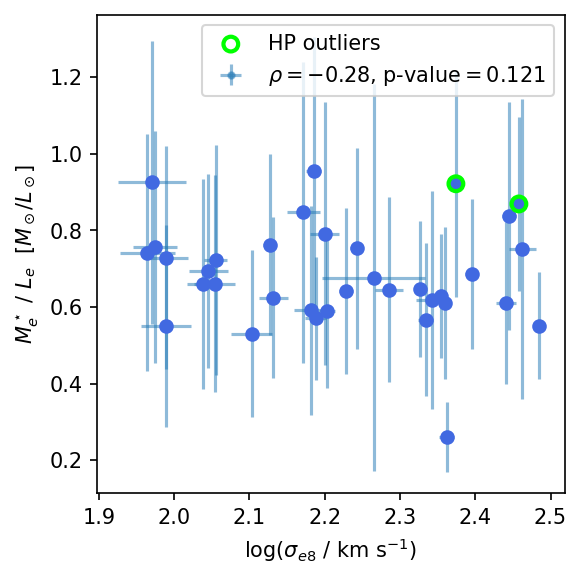}
    \includegraphics[width=0.3\linewidth]{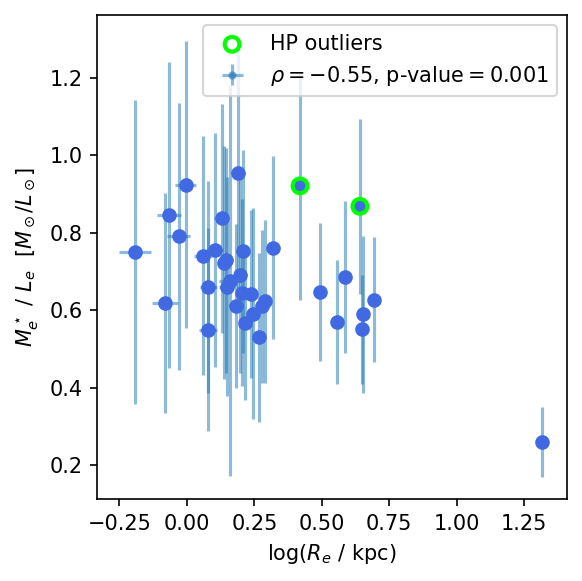}
    \includegraphics[width=0.3\linewidth]{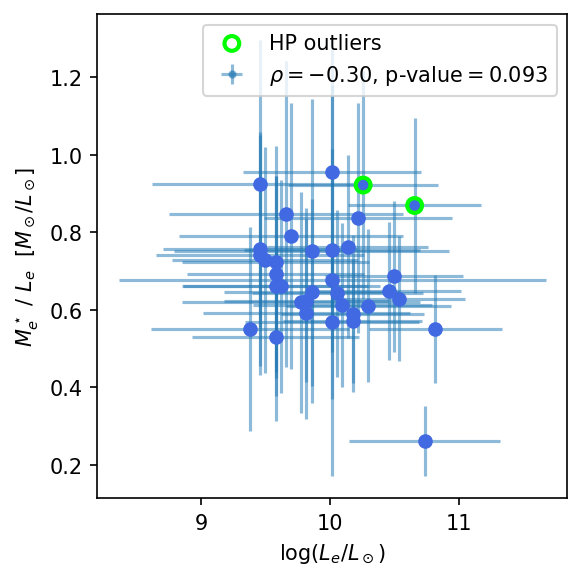}
    \caption{Stellar $M/L$ ratio, computed within the circularized effective radius ($R_{\mathrm{e}}$), as a function of $\log(\sigma_{\mathrm{e8}})$ (left panel), $\log(R_{\mathrm{e}})$ (middle panel), and $\log(L_{\mathrm{e}})$ (right panel). Each plot also reports the Spearman's rank correlation coefficient and its p-value. There is no evident trend in $M_{\mathrm{e}}^\star/L_{\mathrm{e}}$ with respect to $\log(\sigma_{\mathrm{e8}})$ or $\log(L_{\mathrm{e}})$. There seems to be a weak trend with respect to $\log(R_{\mathrm{e}})$ up to $\sim 5.5 \mathrm{kpc}$. The variations in $M_{\mathrm{e}}^\star/L_{\mathrm{e}}$ across this sample of galaxies could explain the remaining scatter in the stellar mass HP.}
    \label{fig:mstar_l}
\end{figure*}

\begin{figure*}
    \centering
    \includegraphics[width=0.45\linewidth]{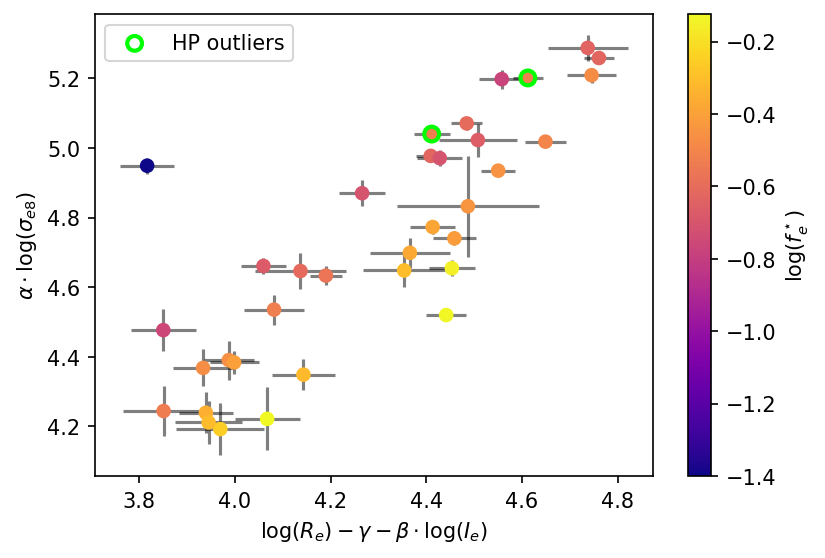}
    \includegraphics[width=0.45\linewidth]{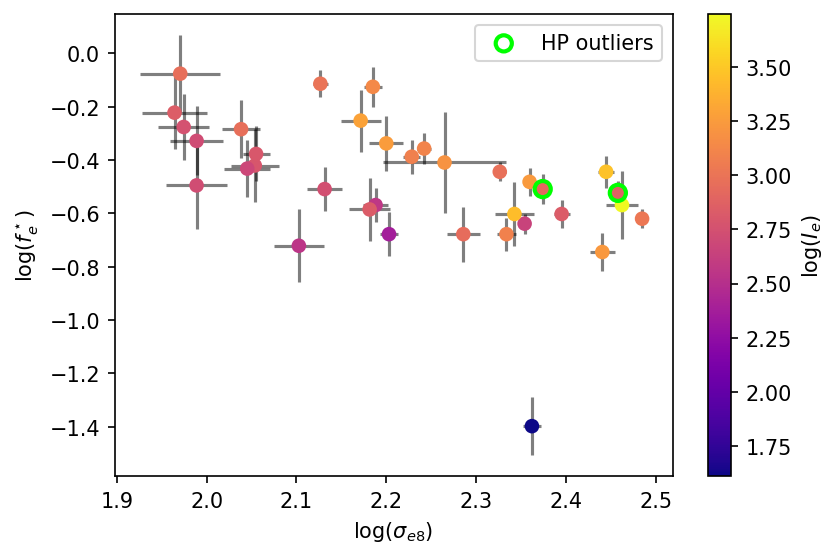}
    \caption{Two projections of the stellar mass fraction HP: one with $\log(\sigma_{\mathrm{e8}})$ as a function of $\log(R_{\mathrm{e}}) - \gamma - \beta \log(I_{\mathrm{e}})$ (left panel) and one with $\log(f_{\mathrm{e}}^{\star})$ as a function of $\log(\sigma_{\mathrm{e8}})$ (right panel). The color of the points indicates values of  $\log(f_{\mathrm{e}}^\star)$ and $\log(I_{\mathrm{e}})$. Both plots highlight the dependence on the stellar mass fraction. The one on the left, in particular, is another way to plot the FP and shows how ETGs with similar stellar mass fractions seem to lie on planes with different slopes and intercepts. The isolated object with a peculiarly low value of $f_{\mathrm{e}}^\star$ is the one with $n = 7.86$ described in Appendix \ref{app:robustness}.}
    \label{fig:fhp_proj}
\end{figure*}

In our sample of ETGs, we do not observe any evident trend in $M_{\mathrm{e}}^\star/L_{\mathrm{e}}$ with respect to $\log(\sigma_{\mathrm{e8}})$ or $\log(L_{\mathrm{e}})$ (see Fig. \ref{fig:mstar_l}), and there appears to be only a weak trend with respect to $\log(R_{\mathrm{e}})$. This, together with the fact that the coefficients $\alpha = 2.0 \pm 0.3$ and $\beta = -1.1 \pm 0.1$ of the $f_{\mathrm{e}}^\star$-HP are consistent with the theoretical values derived by the virial theorem, indicates that the selection criteria we adopted can be used to build a quite pure sample of virialized ETGs, and it corroborates the idea that the major contribution to the tilt of the FP comes from variations in the dark matter content, as highlighted for example in \cite{DeGraaff2022b}, instead of stellar population effects, as suggested for example in \cite{GrilloGobat2010}. The value of the scatter in the $f_{\mathrm{e}}^\star$-HP is $50\%$ lower than that of the classic FP; this confirms what has been observed, for example, by \cite{Dogruel2024}, who find the scatter is $\sim 10 \%$ smaller when extending the FP to a HP. Our sample of galaxies has a mean $\overline{f_{\mathrm{e}}^\star} \approx 37\%$, which corresponds to an average fraction of dark matter within $R_{\mathrm{e}}$ of approximately $\overline{f_{\mathrm{e}}^{DM}} \approx 63\%$. Since $f_{\mathrm{e}}^\star$ is computed using the projected total mass within a cylinder of radius $R_{\mathrm{e}}$, this value cannot be directly compared to those from other studies that estimate the total masses using a 3D dynamical model \citep[e.g.,][]{Gerhard2001,Cappellari2006,Tortora2009,Thomas2007,Thomas2011}; however, it is consistent with the average dark matter fraction value of $\sim 60\%$ measured by \cite{Grillo2010} using 2D total masses and the values found by \cite{DeGraaff2022b} using the EAGLE simulations.

Projecting the $f_{\mathrm{e}}^\star$-HP to lower-dimensional spaces, for example by making an orthogonal projection along the $f_{\mathrm{e}}^\star$ axis, highlights how ETGs with different $f_{\mathrm{e}}^\star$ lie on FPs with different slopes and intercepts (see Fig. \ref{fig:fhp_proj}). Therefore, we can interpret the FP not as a plane in the strict sense, but as a surface that originates from a non-edge-on projection of a higher-dimensional hypersurface, which is in line with the findings of \cite{Yoon2022}.\\

The coefficient $\epsilon = 0.8 \pm 0.2$ of the $f_{\mathrm{e}}^\star$-HP is also compatible with the theoretical value of $1$ within $1\sigma$, even if it is not perfectly on point. This could indicate the presence of measurement errors caused, for example, by $M^\star/L$ gradients, which can bias the estimate of the effective radius \citep{Bernardi2023b}. An alternative explanation is the existence of the so-called weak homology, which implies that the quantity $k_{\mathrm{vir}}$ in Eq. (\ref{eq:fp_virial}) is a function of $R_{\mathrm{e}}$ and $I_{\mathrm{e}}$, while still assuming a constant $M^\star/L$ \citep[see][and references therein]{Bertin2002}. Since we already used $M^\star$ to estimate the stellar mass fraction ($f_{\mathrm{e}}^\star$), only an independent estimation of $M^\star/L$ would reveal the nature of this remaining small tilt. In future works, it could also be worth studying how this coefficient varies by using other mass profiles more complex than the SIS (even at the expense of an increasing number of free parameters).\\

The $M^\star$-HP has a scatter that is half that of the $f_{\mathrm{e}}^\star$-HP, even if the coefficients cannot be linked immediately to any physical-law derivation, like the ones of the $f_{\mathrm{e}}^\star$-HP. However, this is probably why the scatter is so small, since there is no assumption about the shape of the total mass profile or about the $M_{\mathrm{e}}^\star/L_{\mathrm{e}}$ and $f_{\mathrm{e}}^\star$ being constant. Moreover, the $M^\star$ estimates are actually independent of the other three parameters of the HP. Variations in $M_{\mathrm{e}}^\star/L_{\mathrm{e}}$ across this sample of ETGs could instead account for the remaining scatter in the $M^\star$-HP. On the other hand, the fact that the scatter in the $M^\star$-HP is lower than in the $f_{\mathrm{e}}^\star$-HP seems to support the idea that the tilt and the scatter, despite being strictly linked to one another, do not have the same origin. The stellar mass we derived from stellar population analysis could, in fact, also inform about the properties of the stellar population, such as the age and metallicity, that can contribute significantly to the scatter \citep[as shown for example in][]{Gargiulo2009}. In future works, when considering a large sample of galaxies at different redshifts, it will be interesting to determine to what extent the scatter can be further reduced by extending the $M^\star$-HP and $f_{\mathrm{e}}^\star$-HP with other stellar population parameters.

\section*{Data availability}
Table \ref{selected_member_tables} is only available in electronic form at the CDS via anonymous ftp to \url{cdsarc.u-strasbg.fr} (\url{130.79.128.5}) or via \url{ https://cdsarc.cds.unistra.fr/viz-bin/cat/J/A+A/700/A179}

\begin{acknowledgements}
We thank the anonymous referee for their feedback and suggestions that definitely helped to improve the quality of this article.
This work is based on observations taken by the RELICS Treasury Program (GO 14096) with the NASA/ESA HST, which is operated by the Association of Universities for Research in Astronomy, Inc., under NASA contract NAS5-26555.
Based on observations collected at the European Southern Observatory under ESO programme(s) 0102.A-0640(A) and/or data obtained from the ESO Science Archive Facility with DOI(s) under \url{https://doi.org/10.18727/archive/41}.
We acknowledge financial support through grant PRIN-MIUR 2020SKSTHZ. AM acknowledges financial support through grant NextGenerationEU" RFF M4C2 1.1 PRIN 2022 project 2022ZSL4BL INSIGHT.
\end{acknowledgements}

\clearpage

\begin{appendix}

\onecolumn

\section{Tables}
\label{app:tables}

\begin{table}[H]
    \caption{Parameters of the secure ETG sample.}
    \label{selected_member_tables}
    \centering
    \begin{tabular}{c c c c c c c c c c c}
        ID & R.A. & Dec. & $m^{\mathrm{(gal)}}_{\mathrm{F814W}}$ & $n$ & $q$ & $a_{\mathrm{e}}^{(gal)}$ & LOSVD & $M^{\star}$ & $f_{\mathrm{e}}^\star$\\
           & [ICRS] & [ICRS] & [mag.]   &   &   & [$\arcsec$] & [$\mathrm{km}~\mathrm{s}^{-1}$] &  [$10^{10}~\mathrm{M}_{\mathrm{\odot}}$] \\
        \hline\hline
        $266$ & $177.723504$ & $-28.097371$ & $18.3 \pm 0.1$ & $4.10 \pm 0.01$ & $0.710 \pm 0.001$ & $1.010  \pm 0.003$ & $285 \pm 5$ & $14.5 \pm 0.4$ & $0.24 \pm 0.02$ \\
        $344$ & $177.716485$ & $-28.099784$ & $19.1 \pm 0.1$ & $3.82 \pm 0.01$ & $0.790 \pm 0.001$ & $0.826 \pm 0.004$ & $232 \pm 6$ & $8.62 \pm 0.5$ & $0.25 \pm 0.03$ \\
        $366$ & $177.720875$ & $-28.098379$ & $19.2 \pm 0.1$ & $3.16 \pm 0.01$ & $0.947 \pm 0.001$ & $0.610 \pm 0.002$ & $198 \pm 4$ & $7.4 \pm 0.2$ & $0.36 \pm 0.03$ \\
        $430$ & $177.728033$ & $-28.095713$ & $19.7 \pm 0.1$ & $4.21 \pm 0.02$ & $0.791 \pm 0.001$ & $0.564 \pm 0.004$ & $221 \pm 5$ & $6.7 \pm 0.5$ & $0.31 \pm 0.04$ \\
        $\vdots$ & $\vdots$ & $\vdots$ & $\vdots$ & $\vdots$ & $\vdots$ & $\vdots$  & $\vdots$ & $\vdots$ & $\vdots$ \\
        \hline
    \end{tabular}
\end{table}
\tablefoot{This is an extract of the full table, which is only available in electronic form at the CDS. The first column (ID) contains the ID of the objects from the spectroscopic catalog published in \citetalias{DAddona2024}; the second and third columns show the J2000 ICRS right ascension and declination (R.A and Dec.); the fourth one ($m_{\mathrm{F814W}}^{\mathrm{(gal)}}$) contains the magnitude from the Galfit best-fit made in HST band F814W; the fifth column ($n$) reports the S\'ersic index;  the sixth one ($q$) contains the axis ratio; the seventh one ($a_{\mathrm{e}}^{\mathrm{(gal)}}$) contains the Galfit effective semimajor axis in arcseconds; the eighth column reports the LOSVD in $\mathrm{km}~\mathrm{s}^{-1}$; the ninth one ($M^\star$) contains the total stellar mass in units of $10^{10}~\mathrm{M}_{\mathrm{\odot}}$; the tenth one ($f_{\mathrm{e}}^\star$) contains the stellar-over-total mass fraction, namely, the ratio between the stellar mass and the total mass within a circle of radius $R_{\mathrm{e}}$. In the electronic version of this table, the errors on the parameters are reported in columns eleven to seventeen.}

\section{Additional figures}
\label{app:additional_figures}

\begin{figure*}[h!]
    \centering
    \resizebox{14cm}{14cm}{%
        \includegraphics[width=\linewidth]{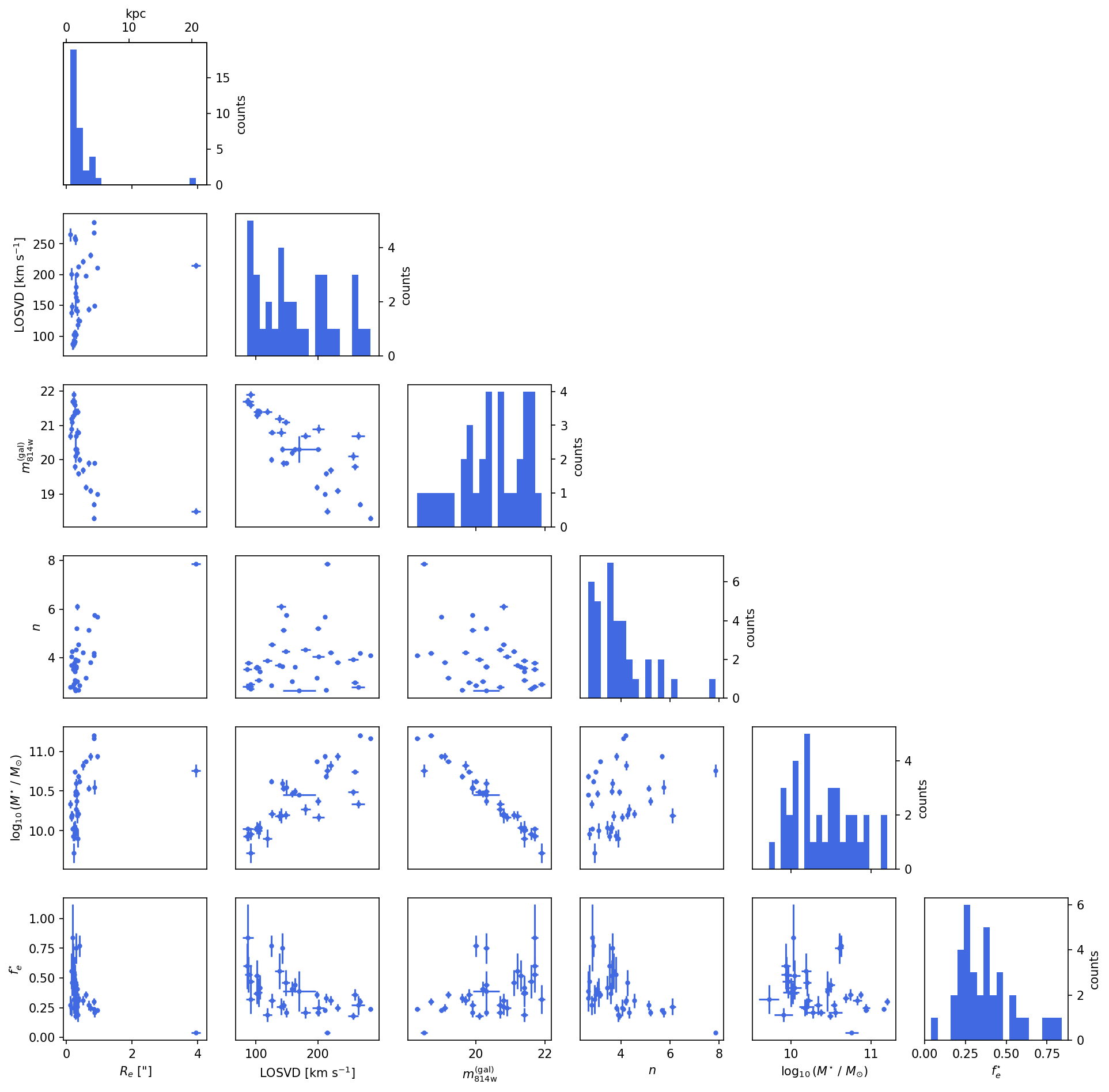}
    }
    \caption{Distribution of the ETGs used to fit the FP as a corner plot of: the galfit magnitude in HST band F814W, the LOSVD, the circularized effective radius ($R_{\mathrm{r}}$), the S\'ersic index, the stellar mass, and the stellar-over-total mass fraction ($f_{\mathrm{e}}^{\star}$).}
    \label{fig:hists}
\end{figure*}

\clearpage

\section{Robustness of hyperplane fitting}
\label{app:robustness}
\begin{figure*}[h]
    \centering
    \includegraphics[width=0.75\linewidth]{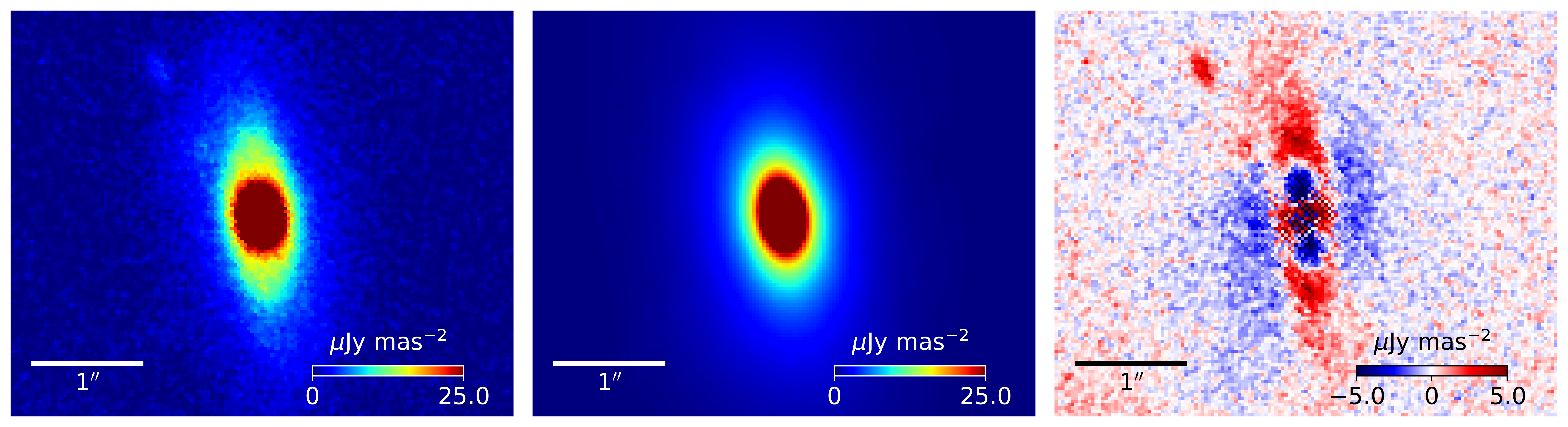}
    \includegraphics[width=0.75\linewidth]{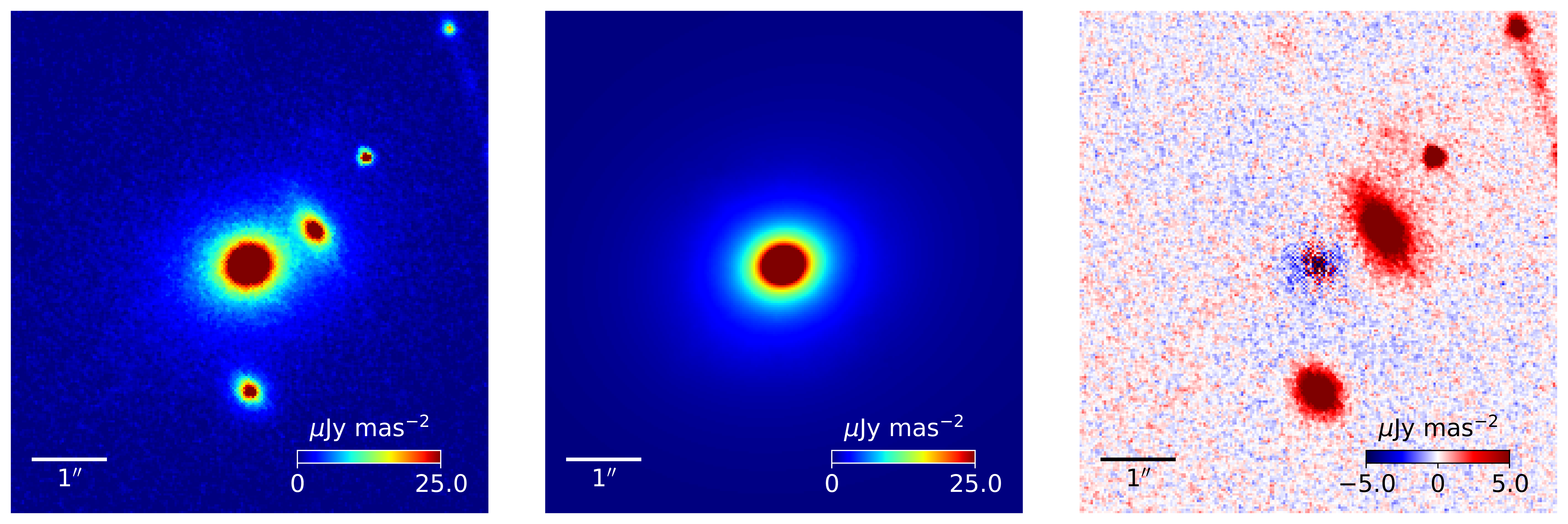}
    \caption{Upper panel: One of the three galaxies that were removed from the original sample due to the presence of substructures in the residuals of the Morphofit model. Lower panel: Galaxy with $n=7.86$ that has been kept in the sample. From left to right, the columns show a cutout of the object taken from the HST F814W image, the corresponding Morphofit model, and the residual image.}
    \label{fig:big_sersic}
\end{figure*}

Initially, our sample of ETGs contained 38 galaxies, 4 of which had a suspiciously large S\'ersic index value. For this reason, we re-inspected the residuals of the Morphofit models and found that three of them actually showed faint substructures (see Fig. \ref{fig:big_sersic}, upper panel). Therefore, these three objects were then excluded from the sample, according to the selection criteria we illustrated in Sect. \ref{sec:analysis}. The fourth one (ID=1129), with $n=7.86$, has been kept in the sample as it did not present any sign of substructures, or an evident issue in the Morphofit model (Fig. \ref{fig:big_sersic}, lower panel). Interestingly, the value of the coefficients of the FP and the two HPs did not change significantly when the fit was performed using the sample of ETGs containing these three objects. This suggests that the fitting procedure is quite robust against the presence of spurious objects that do not strictly follow our selection criteria.

\begin{figure*}[h]
    \centering
    \includegraphics[width=0.45\linewidth]{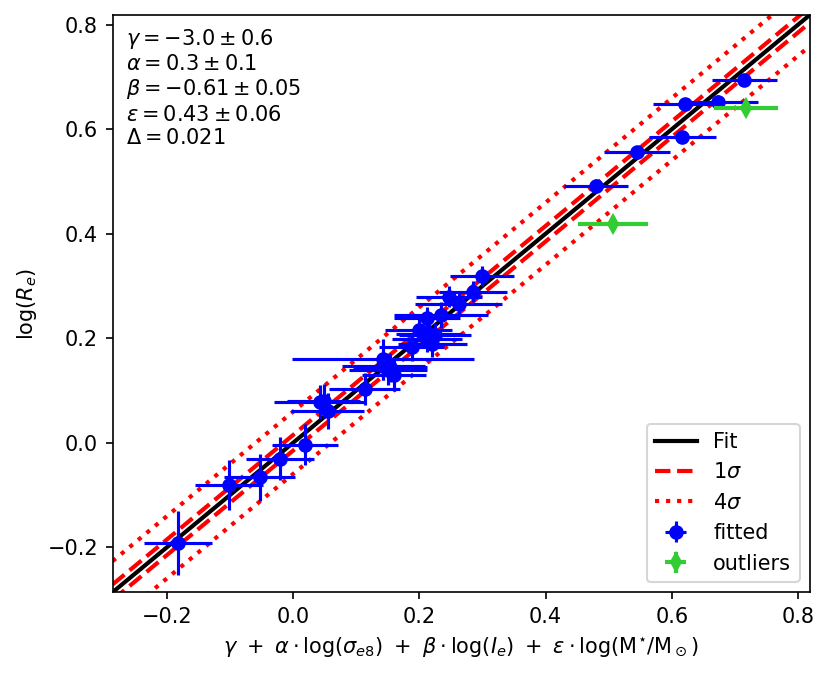}
    \includegraphics[width=0.45\linewidth]{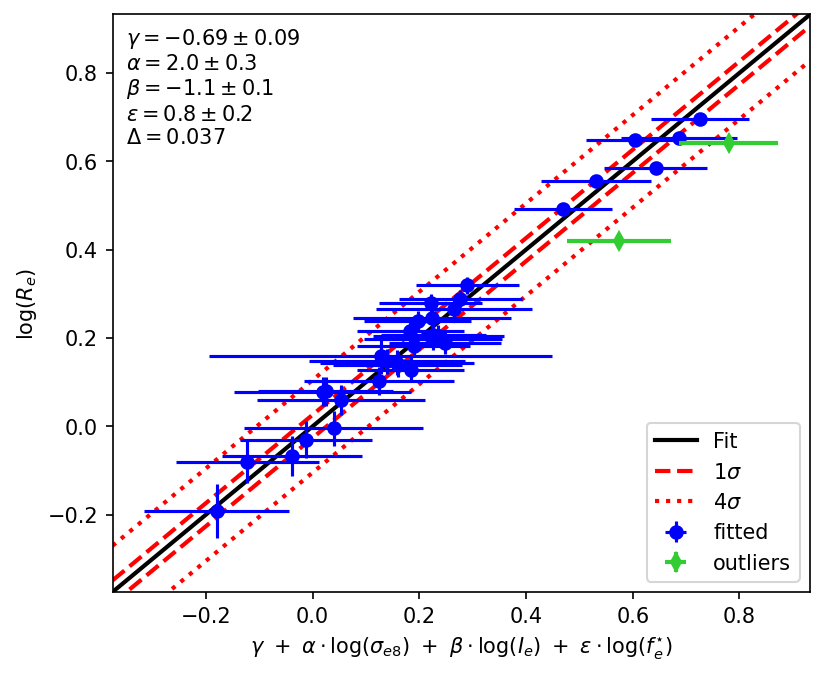}
    \caption{Fit to the stellar mass and mass fraction HPs performed excluding the object with $n=7.86$. The solid black line is the projection of the best-fitting plane, and the dashed and dotted red lines indicate the $1$ and $4$ standard deviation intervals, respectively. Green points are objects identified as outliers and therefore not included in the fit.}
    \label{fig:non_fhp}
\end{figure*}

The fourth object with ID=1129, due to its particularly large $R_{\mathrm{e}}$, lies on the top right corner of the FP and HP plots. One could argue that, since it is relatively far from the other galaxies in the parameter space, it could have a large influence on the slope of the best fitting planes. So, to check whether this is the case, we performed the fit excluding this object and found that also in this case, the value of the coefficients of the two HPs are consistent with ones we previously found (see Fig. \ref{fig:non_fhp}). This not only corroborates the robustness of the fitting procedure, but it also confirms that this object follows the FP/HP scaling relations.

\end{appendix}

\end{document}